\documentclass[preprint,12pt]{./elsarticle}

\usepackage{amssymb}

\usepackage{amsfonts,amsmath,amssymb,amsbsy,amsthm}

\usepackage{hyperref}
\usepackage{mathrsfs}
\usepackage{epstopdf}

\usepackage{graphicx}

\usepackage{subcaption}

\usepackage{bbm} 

\begin{document}

\begin{frontmatter}

\title{A Critical Study of Howell \emph{et al.}'s Nonlinear Beam Theory}

\author{Kavinda Jayawardana\corref{mycorrespondingauthor}}
\cortext[mycorrespondingauthor]{Corresponding author}
\ead{zcahe58@ucl.ac.uk}

\begin{abstract}
In our analysis, we show that Howell \emph{et al.}'s nonlinear beam theory \cite{howell2009applied} does not depict a representation of the Euler-Bernoulli beam equation, nonlinear or otherwise. The authors' nonlinear beam theory implies that one can bend a beam in to a constant radius of deformation and maintain that constant radius of deformation with zero force. Thus, the model is disproven by showing that it is invalid when the curvature of deformation is constant, while even the linear Euler-Bernoulli beam equation stays perfectly valid under such deformations. To conclude, we derive a nonlinear beam equation by using Ciarlet's nonlinear plate equations \cite{Ciarlet1997theory} and show that our model is valid for constant radius of deformations.
\end{abstract}


\end{frontmatter}

\section{Introduction}
\label{S:1}

A \emph{beam} is a structural element with one planar dimension is larger in comparison to its thickness and its other planar dimension. Thus, beam theories are derived  from the three-dimensional elastic theory by making suitable assumptions concerning the kinematics of deformation or the state stress through the thickness, thereby reducing three-dimensional elasticity problem into a one-dimensional problem. Such theories provide methods of calculating the load-carrying and deflection characteristics of beams, where the most notable beam theory is the Euler-Bernoulli beam theory \cite{timoshenko1953history}: for the zero-Poisson's ratio case, it has a governing equation of the following form,
\begin{align}
EI \frac{\partial^4 w(x)}{\partial x^4} - q(x) =0~, \label{beam}
\end{align}
where $w(x)\in \mathbb R$ is the deflection (i.e. the displacement or the deformation form the $z=0$ line, with respect to the 3D Cartesian coordinate system), $E$ is the Young's modulus, $I$ is the second moment of inertia of the beam and $q(x)$ is a transverse load acting on the beam.\\

Euler-Bernoulli beam theory is a good approximation for the case of small deflections of a beam that are subjected to lateral loads. However, if one also takes into account the shear deformation and the rotational inertia effects, i.e. making it suitable for describing the behaviour of thick beams, then the most acceptable beam theory is considered to be is the Timoshenko beam theory \cite{timoshenko1922x}. In fact, Euler-Bernoulli beam theory is a special case of Timoshenko beam theory. Note that the governing equations for Timoshenko beam theory can be derived  by eliminating the $y$ dependency from Mindlin-Reissner plate theory.\\

Another approach to study the nonlinear behaviour of beams was attempted by Howell, Kozyreff and Ockendon \cite{howell2009applied} in their publication \emph{Applied Solid Mechanics}\footnote{\href{https://books.google.co.uk/books?id=WoGQmI_vZMAC\&pg=PA187\&lpg=PA187\&dq=Applied+Solid+Mechanics+Nonlinear+beam+theory+Howell\&source=bl\&ots=lTtkl662fD\&sig=kHfgYh67lciNE8k_5a6Pth0Szn8\&hl=en\&sa=X\&ved=0ahUKEwjKsKTy7aLMAhUDWRoKHRWaBcwQ6AEINjAD\#v=onepage\&q=Applied\%20Solid\%20Mechanics\%20Nonlinear\%20beam\%20theory\%20Howell\&f=false}{ https://books.google.co.uk/books?isbn=0521671094}}, where the authors claim that their formulation is an alternative representation of the nonlinear Euler-Bernoulli beam theory (see page 190 of Howell \emph{et al.} \cite{howell2009applied}). In our reading, we show that the authors' model is flawed and it does not depict a representation of a beam equation, nonlinear or otherwise, as it is invalid for constant radius of deformations. To conclude we derive a nonlinear beam equation by the use of Ciarlet's nonlinear plate equations \cite{Ciarlet1997theory} (see section B of Ciarlet \cite{Ciarlet1997theory}) with mathematical regiour, and show that our model is valid for constant radius of deformations.

\section{Howell \emph{et al.}'s Nonlinear Beam Theory}

\begin{figure}[!h]
\centering
\includegraphics[width=0.75\linewidth]{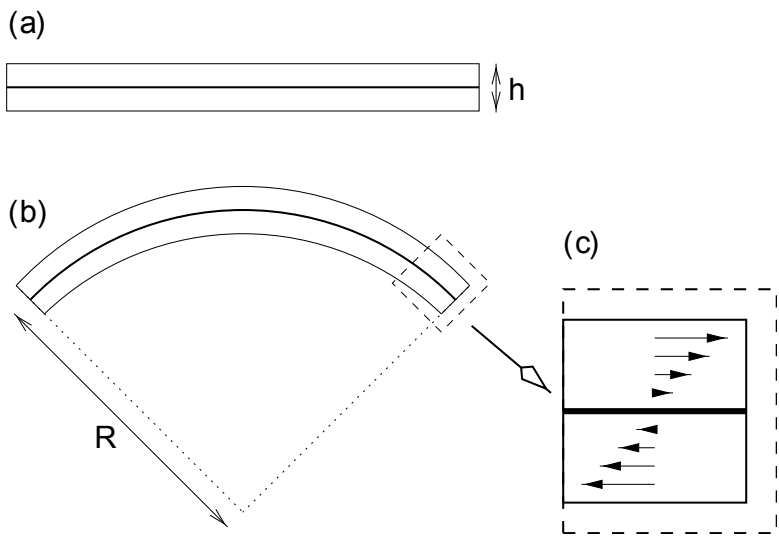}
\caption{`A beam (a) before bending and (b) after bending; (c) a close-up of the displacement field' \cite{howell2009applied}, where $h$ is the thickness of the beam and $R$ is the radius of the curvature of deformation. \label{HowellWrong}}
\end{figure}

In section 4.3. of Howell \emph{et al.} \cite{howell2009applied}, the authors present a method for studying the behaviour of large deflections of beams. With methods that are inconsistent with the finite-deformation theory and differential geometry, the authors derive their own version of the nonlinear Euler-Bernoulli beam equation, where the authors define the governing equation to this model as follows,
\begin{align}
EI \frac{d^2 \theta(s)}{ds^2} + N_0\cos\theta-T_0\sin\theta = 0 ~, \label{Howell}
\end{align}
where $\theta(s)$ and $s$ are respectively the angle and the arc length between the centre-line of the beam (see figure \ref{HowellWrong}), and $T_0$ and $N_0$ are the forces applied at the boundary which are parallel to the $x$ and $y$ axis respectively (note that the authors' $y$ dimension is the $z$ or the transverse-dimension in terms of the standard beam theory terminology). The authors assume that the centre-line is \emph{virtually unchanged}, and thus, large in-plane strains of the centre-line are ignored in the derivation of the equation (\ref{Howell}). However, what the authors put forward is flawed and it does not depict a representation of the Euler-Bernoulli beam equation, nonlinear or otherwise. For a comprehensive study of the nonlinear Euler-Bernoulli beam equation please, consult Reddy \cite{reddy2014introduction} or Hodges \emph{et al.} \cite{hodges1980nonlinear}.\\

To examine why equation (\ref{Howell}) is flawed, consider a case where one is bending a beam into a shape with constant radius of curvature $R$ (i.e. $\theta(s) = R^{-1}s$), where $R^3\gg I$, by applying appropriate boundary forces $N_0$ and $T_0$. Then equation (\ref{Howell}) reduces to $$EI\frac{d^2 }{ds^2}\left(\frac{s}{R}\right) + N_0\cos\theta-T_0\sin\theta = 0~.$$ As $R$ is a constant, one finds that $ N_0\cos\theta-T_0\sin\theta = 0$, $\forall~\theta$, and thus, $ N_0=0$ and $ T_0 = 0$. This implies that it takes zero force to bend the beam with a constant radius of deformation, regardless of the magnitude of the deformation, which is not physically viable. Now, the reader can see that when the radius of curvature of deformation is constant, equation (\ref{Howell}) is no longer valid.\\

However, those who are familiar with beam theories may argue that one cannot bend a beam so that the radius of the deformation is constant without an external forcing. Thus, if one considers equation 4.9.3 of Howell \emph{et al.} \cite{howell2009applied}, then one can re-express equation (\ref{Howell}) as follows,
\begin{align}
EI \frac{d^2 \theta(s)}{ds^2} + N(\theta)= 0~.\label{Howell2}
\end{align}
where $N(\theta)$ is an normal force acting on the beam. Again, bend the beam in to a constant radius of curvature $R$ by applying a normal force of $N(\theta)$ to obtain $EI\frac{d^2 }{ds^2}(R^{-1}s) + N(\theta)= 0$, which implies that $N(\theta)=0$, $\forall \theta$. As the reader can see that the authors' nonlinear beam equation is, again, no longer valid, regardless of the magnitude of the deformation, i.e. regardless of the limits of $\theta$ or the magnitude of the constant $R$.\\

On another note, equation (\ref{Howell}) implies that one can bend a beam in to a constant radius of deformation and maintain that constant radius of deformation with zero force. To illustrate this, consider the deformation of a diving board as a thought experiment (to stay in consistent with the numerical modelling of section 4.9.2 of Howell \emph{et al.} \cite{howell2009applied}). If one bends this diving board to either semicircular shape with radius $R$ or over a barrel with a radius $R$ respectively, then equation (\ref{Howell}) or equation (\ref{Howell2}) imply that $N_0, T_0=0$ (zero boundary forces) or $N(\theta)=0$ (zero transverse load), $\forall \theta$, respectively, as now $EI \frac{d^2 \theta}{ds^2} \equiv 0$. Can the reader see why this is not physically realistic? \\

In contrast, consider that if one were to bend a beam into a shape with a constant radius $R$ by a transverse load $q(x)$ below the $x$-axis (i.e. below $z=0$ line) with respect the linear Euler-Bernoulli beam equation (\ref{beam}), then one finds that
\begin{align}
EI\left[3\frac{1}{(u(x)+R)^{3}}+18\frac{x^2}{(u(x)+R)^{5}}+15\frac{x^4}{(u(x)+R)^{7}}\right] + q(x)=0~, \label{beam2}
\end{align}
where $u(x) = \sqrt{R^2\!-\!x^2}-R$ and $x\ll R$. As the reader can see that Euler-Bernoulli equation stays perfectly valid under such deformations. Equation (\ref{beam2}) also shows that (i.e. Euler-Bernoulli equation implies that) it takes a non-zero transverse load (i.e. $q(x)\neq0$) to bend a beam in to a constant radius of curvature, which directly contradicts Howell \emph{et al.}'s model \cite{howell2009applied}. Can the reader see now why Howell \emph{et al.}'s  beam theory  \cite{howell2009applied} is flawed?\\

The flaws of equation (\ref{Howell}) has arisen from the derivation of the equation (see section 4.9.1 of Howell \emph{et al.} \cite{howell2009applied}) as the authors did not consider any valid mathematical techniques in the study of thin objects subjected to finite-strains and mathematical techniques in study of coordinate transforms. The authors make the fundamental mistake of assuming that an arbitrary coordinate transform is the same as a deformation of an elastic body (see equations 4.9.1 of of Howell \emph{et al.} \cite{howell2009applied}), resulting in a nonsensical model that is inconsistent with the behaviour of beams.\\

To conclude our analysis, we derive a nonlinear beam equation by using Ciarlet's nonlinear plate equations \cite{Ciarlet1997theory} (see section B of Ciarlet \cite{Ciarlet1997theory}). For the boundary conditions, as one needs to consider the \emph{pure-traction} case and the \emph{displacement-traction} case separately, we only consider the pure-traction case as this appears to be the only case that is consistent Howell \emph{et al.}'s derivation \cite{howell2009applied}. Note that \emph{Einstein's summation notation} is assumed  throughout, we regard the indices $i,j,k,l \in \{1,2,3\}$ and  $\alpha,\beta,\gamma,\delta \in \{1,2\}$, and the coordinates $(x^1,x^2,x^3)=(x,y,z)$, unless it is strictly states otherwise.\\

Recall Ciarlet's nonlinear plate equations \cite{Ciarlet1997theory}, and thus, one can express the energy functional of a plate for the pure-traction case as follows,
\begin{align}
J(\boldsymbol u) = h\int\limits_\Omega \Bigg[\frac{1}{8} A^{\alpha \beta \gamma \delta} & \left(\partial_{\alpha} \bar R_{ i}(\boldsymbol u) \partial_{\beta} \bar R^{ i}(\boldsymbol u) -\delta_{ \alpha \beta}\right) \left(\partial_{ \gamma} \bar R_{k}(\boldsymbol u) \partial_{\delta} \bar R^{ k}(\boldsymbol u) -\delta_{\gamma\delta}\right) \nonumber \\
+\frac{1}{24}h^2A^{\alpha  \beta  \gamma  \delta} & \left(\partial_{ \alpha  \beta} \bar R_{ i}(\boldsymbol u) \frac{\left(\partial_x \bar{\boldsymbol R}(\boldsymbol u) \times \partial_y \bar{\boldsymbol R}(\boldsymbol u)\right)^{ i}}{|| \partial_x \bar{\boldsymbol R} (\boldsymbol u) \times \partial_y \bar{\boldsymbol R}(\boldsymbol u)||}\right ) \nonumber \\
 & \left(\partial_{ \gamma  \delta} \bar R_{ k}(\boldsymbol u) \frac{\left(\partial_x \bar{\boldsymbol R} (\boldsymbol u) \times \partial_y \bar{\boldsymbol R}(\boldsymbol u)\right)^{ k}}{|| \partial_x \bar{\boldsymbol R} (\boldsymbol u) \times \partial_y \bar{\boldsymbol R}(\boldsymbol u)||}\right )   -  f^iu_i\Bigg]dx dy~,\nonumber\\
&-h\int\limits_{\partial\Omega} \Bigg[\tau_0^i u_i + \frac{1}{12}h^2\eta_0^{i\alpha}\partial_\alpha u_i \Bigg]d(\partial\Omega) \\
\boldsymbol u  \in & \{ \boldsymbol v \mid \partial_{\alpha} \bar R_{ i}(\boldsymbol v) \partial_{\beta} \bar R^{ i}(\boldsymbol v) \neq \boldsymbol 0 ~,~\forall \boldsymbol(x,y\boldsymbol) \in \Omega \}~, \nonumber
\end{align}
where 
\begin{align*}
\bar{\boldsymbol R}(\boldsymbol u) = \boldsymbol ( x,y,0 \boldsymbol) + \boldsymbol(u^{1}(x,y),u^{2}(x,y),u^{3}(x,y) \boldsymbol)~,
\end{align*}
and where $\boldsymbol u$ is the displacement field, $\boldsymbol f$ is an external force density field, $\boldsymbol \tau_0$ is an external traction (i.e. applied stress) field, $\boldsymbol \eta_0$ is an external change in moments density (i.e. applied change in torque density) field,
\begin{align*}
A^{\alpha\beta\gamma\delta} = \frac{2\mu\lambda}{(\lambda+2\mu)}\delta^{\alpha\beta}\delta^{\gamma\delta} + \mu\left( \delta^{\alpha\gamma}\delta^{\beta\delta}+\delta^{\alpha\delta}\delta^{\beta\gamma}\right)
\end{align*}
is the elasticity tensor for a plate, $\partial_i$ is the partial derivate with respect to the coordinate $x^i$, $\delta^j_i$ is the Kronecker delta, $\times$ is the Euclidean cross-product, $||\cdot||$ is the standard Euclidean norm, $\Omega \subset \mathbb{R}^2$ is a 2D-plane that describes the unstrained mid-plane of the plate. Note that 
\begin{align*}
\lambda &=\frac{\nu E}{(1+\nu)(1 -2\nu)}~\text{and}\\
\mu &= \frac{E}{2(1+\nu)}
\end{align*}
are the first and the second Lam\'{e}'s parameters of the plate respectively, and $\nu$ is the
Poisson ratio and $E$ Young's modulus of the plate.\\

Now, assume that one is considering a plate with infinite length in $y$-dimension, and, also assume that the plate is independent of any displacements in the $x$-dimension, and thus, we may assume that $u^1=0$ and $u^2=0$. Thus, the energy functional of a plate per unit $y$ can be expressed as follows,
\begin{align*}
J(\boldsymbol u) = h\int\limits_\omega \Bigg[\frac{1}{8} A^{1111} & \left(\partial_{1} \bar R_{ i}(\boldsymbol u) \partial_{1} \bar R^{ i}(\boldsymbol u) -\delta_{11}\right) \left(\partial_{ 1} \bar R_{k}(\boldsymbol u) \partial_{1} \bar R^{ k}(\boldsymbol u) -\delta_{11}\right) \nonumber \\
+\frac{1}{24}h^2A^{1111} & \left(\partial_{11} \bar R_{ i}(\boldsymbol u) \frac{\left(\partial_x \bar{\boldsymbol R}(\boldsymbol u) \times \partial_y \bar{\boldsymbol R}(\boldsymbol u)\right)^{ i}}{|| \partial_x \bar{\boldsymbol R} (\boldsymbol u) \times \partial_y \bar{\boldsymbol R}(\boldsymbol u)||}\right )\\
& \left(\partial_{ 11} \bar R_{ k}(\boldsymbol u) \frac{\left(\partial_x \bar{\boldsymbol R}(\boldsymbol u) \times \partial_y \bar{\boldsymbol R}(\boldsymbol u)\right)^{ k}}{|| \partial_x \bar{\boldsymbol R} (\boldsymbol u) \times \partial_y \bar{\boldsymbol R}(\boldsymbol u)||}\right )   -  f^3u_3\Bigg]dx \\ 
&-h\Bigg[\tau_0^3 u_3+ \frac{1}{12}h^2\eta_0^{31}\partial_1 u_3 \Bigg]|_{\partial\omega} \\
\boldsymbol u  \in & \{\boldsymbol v  \mid \partial_{1} \bar R_{ i}(\boldsymbol v) \partial_{1} \bar R^{ i}(\boldsymbol v) \neq \boldsymbol 0 ~,~\forall x \in \omega \}~,
\end{align*}
where $\omega \subset \mathbb{R}$ describes the unstrained centre-line of the beam. Now, to stay consistent with the standard beam theory notations, we redefine $u^3 =w$, $f^3= \tilde q ~(=q/h)$. We also let $\tau^3 = \tau_0$ and $\eta_0^{31} = \eta_0$ for convenience, were $\tau_0$ is an external normal traction and $\eta_0$ is an external change in normal moments density (note that $\tau_0$ and $\eta_0$ are not constants as they may attain different values on the boundary of the set $\omega$, i.e. in $\partial\omega$). Thus, one can express an energy functional of a nonlinear beam as follows,
\begin{align}
J(w) = h\int\limits_\omega  \Bigg[\frac{1}{8}  \Lambda (\partial_{x} w )^4 &+\frac{1}{24}h^2 \Lambda \frac{ (\partial_{xx} w )^2}{1 + (\partial_x w)^2}    -  \tilde q w\Bigg]dx \nonumber \\
&-h \Bigg[\tau_0 w + \frac{1}{12}h^2\eta_0 \partial_x w \Bigg]|_{\partial\omega}~, \label{nonlinearfunc}\\
w  \in & \{ v  \mid \partial_x v \neq 0 ~,~\forall x \in \omega \}~  ,\nonumber
\end{align}
where 
\begin{align*}
\Lambda &=4\mu\frac{(\lambda+\mu)}{(\lambda+2\mu)}\\
& = \frac{E}{(1+\nu)(1 -\nu)}~.
\end{align*}
To derive the governing equations, we apply \emph{principle of virtual displacements} (see section 2.2.2 of Reddy \cite{reddy2006theory}) to equation (\ref{nonlinearfunc}) to obtain the following,
\begin{align*}
J'(w)(\delta w) = h\int\limits_\omega & \Bigg[ \frac{1}{2}  \Lambda  (\partial_{x} w )^3 \partial_{x} \delta w  \\
& +\frac{1}{12}h^2 \Lambda  \Big[ \frac{ \partial_{xx} w }{1 + (\partial_x w)^2}\partial_{xx}\delta w -\frac{ (\partial_{xx}w )^2 \partial_x w}{(1 + (\partial_x w)^2)^2}\partial_{x} \delta w \Big]  - \tilde q \delta w\Bigg]dx \nonumber\\
&-h \Bigg[\tau_0 \delta w + \frac{1}{12}h^2\eta_0 \partial_x \delta w \Bigg]|_{\partial\omega} ~,\\
w  \in & \{ v  \mid \partial_x v \neq 0 ~,~\forall x \in \omega \}~  .
\end{align*}
Now, applying integration by parts and collecting all the $\delta w$ terms to obtain a \emph{nonlinear beam equation} for a traverse loading $q(x)$ of the following form,
\begin{align}
\frac{1}{12}h^2 \Lambda \Bigg[ \partial_{xx}\left(\frac{ \partial_{xx} w }{1 + (\partial_x w)^2}\right) + \partial_x \left(\frac{ (\partial_{xx} w)^2 \partial_x w}{(1 + (\partial_x w)^2)^2} \right)\Bigg]& \nonumber\\
 - \frac{1}{2} \Lambda \partial_{x} (\partial_{x} w )^3 - \tilde q(x)  & = 0 ~,\label{nonlinearbeam}
\end{align}
where the \emph{boundary conditions} can be expressed as follows,
\begin{align*}
\Bigg [\frac{1}{2} \Lambda  (\partial_{x} w )^3 - \frac{1}{12}h^2 \Lambda \Big[ \partial_{x}\left(\frac{ \partial_{xx} w }{1 + (\partial_x w)^2}\right) + \frac{ (\partial_{xx} w)^2 \partial_x w}{(1 + (\partial_x w)^2)^2} \Big] \Bigg]|_{\partial\omega} & = \tau_0 ~,\\
 \Lambda \Bigg[ \frac{ \partial_{xx} w }{1 + (\partial_x w)^2} \Bigg]|_{\partial\omega} &= \eta_0 ~.
\end{align*}

To emphasise to the reader that equation (\ref{nonlinearbeam}) does, indeed, represents the governing equations of a beam, should one linearise equation (\ref{nonlinearbeam}), one gets the following,
\begin{align*}
\frac{1}{12}h^2 \Lambda \frac{\partial^4 w(x)}{\partial x^4} - \tilde q(x)  = 0 ~,
\end{align*}
which is the Euler-Bernoulli beam equation for a non-zero Poisson's ratio, where the linearised boundary conditions can be expressed as follows,
\begin{align*}
- \frac{1}{12}h^2 \Lambda  \frac{\partial^3 w(x)}{\partial x^3} |_{\partial\omega} & = \tau_0 ~,\\
 \Lambda  \frac{\partial^2 w(x)}{\partial x^2}  |_{\partial\omega} &= \eta_0 ~.
\end{align*}

Now, recall Howell \emph{et al.}'s equations 4.9.1  \cite{howell2009applied} (see page 118 of Howell \emph{et al.} \cite{howell2009applied}), i.e.
\begin{align}
\frac{\partial x(s)}{\partial s}  & = \cos(\theta(s)) ~~\text{and} \label{eq1}\\
\frac{\partial w(s)}{\partial s}  & = \sin(\theta(s)) ~. \label{eq2}
\end{align}
Note that in our analysis, we do not treat equations (\ref{eq1}) and (\ref{eq2}) as coordinate transforms or as deformations, as equations (\ref{eq1}) and (\ref{eq2}) satisfies neither definitions. We merely treat  these equations as change of variables. Thus, by the changing variables in equation (\ref{nonlinearbeam})  with equations (\ref{eq1}) and (\ref{eq2}), one finds the following,
\begin{align}
\frac{1}{12}h^2 \Lambda \sec^3(\theta) \Bigg[ \frac{\partial^3 \theta(s)}{\partial s^3} + 6 \tan(\theta)\frac{\partial \theta(s)}{\partial s} \frac{\partial^2 \theta(s)}{\partial s^2} & \nonumber\\
+ 4\sec^2(\theta) \left(\frac{\partial \theta(s)}{\partial s} \right)^3 + 2 \left(\frac{\partial \theta(s)}{\partial s} \right)^3 & \Bigg] \nonumber\\
- \frac{3}{2} \Lambda \sec^3(\theta)\tan^2(\theta) \frac{\partial \theta(s)}{\partial s} - &\tilde  q(s)  = 0 ~,\label{goodstuff}
\end{align}
where the boundary conditions can be expressed as follows,
\begin{align*}
\Bigg[\frac{1}{2} \Lambda \tan^3(\theta)  - \frac{1}{12}h^2 \Lambda \sec^2(\theta) \Big[ \frac{\partial^2 \theta(s)}{\partial s^2} + 2 \tan(\theta)\left(\frac{\partial \theta(s)}{\partial s} \right)^2 \Big] \Bigg]|_{\partial\omega} &=  \tau_0 ~,\\
\Big[\Lambda \sec (\theta)\frac{\partial \theta(s)}{\partial s} \Big]|_{\partial\omega} & = \eta_0~.
\end{align*}

Recall that what Howell \emph{et al.} \cite{howell2009applied} present as the nonlinear beam equation fails to be valid under constant curvature of deformation. Thus, consider the case where one bends a beam in to a constant radius of curvature, where the radius is $r$, i.e. $\theta(s) = s/r$. Thus, equation (\ref{goodstuff}) implies a transverse load of the form, 
\begin{align}
\tilde q(s) = & \frac{1}{6}\left(\frac{h}{r}\right)^2 \frac{\Lambda}{r} \sec^3(s/r) \Big[ 2\sec^2(s/r) + 1 \Big] \nonumber\\
&- \frac{3}{2} \frac{\Lambda}{r} \sec^3(s/r)\tan^2(s/r)~,\label{const}
\end{align}
and where the boundary conditions can be expressed as follows,
\begin{align*}
\Bigg[\frac{1}{2} \Lambda \tan^3(s/r)  - \frac{1}{6}\left(\frac{h}{r}\right)^2 \Lambda \sec^2(s/r)\tan(s/r) \Bigg]|_{\partial\omega} &=  \tau_0 ~,\\
\Bigg[\frac{\Lambda}{r} \sec (s/r) \Bigg]|_{\partial\omega} & = \eta_0~.
\end{align*}

From equation (\ref{const}), one can see that a non-zero transverse load is required to deform a beam in to a constant radius of curvature, and the magnitude (i.e. the strength of the force required) is a function of the curvature of deformation (i.e. $1/r$) and the limits of the arc $s$, which far more realistic than what Howell \emph{et al.} \cite{howell2009applied} present.

\section{Conclusions}

With a counter example, we showed that Howell \emph{et al.}'s nonlinear beam theory \cite{howell2009applied} is flawed. The model was disproven by showing that it is invalid when the curvature of deformation is constant, while even the linear Euler-Bernoulli beam equation stays perfectly valid under such deformations. Should the reader examine section 4.9.1 of Howell \emph{et al.} \cite{howell2009applied}, it becomes evident that the authors assumes an arbitrary coordinate transform is the same as a deformation of an elastic body. To understand the distinction between a coordinate transform and a vector displacement, we refer the reader to   section 1 and section 2 of Morassi and Paroni \cite{Morassi}, and for a comprehensive mathematical study of thin bodies (plates and shells) subjected to finite-strains, we refer the reader to sections Bs of Ciarlet \cite{Ciarlet1997theory} and Ciarlet \cite{Ciarlet}.\\

With our analysis, we also showed that that Howell \emph{et al.}'s nonlinear beam theory \cite{howell2009applied} implies that one can bend a beam in to a constant radius of deformation and maintain that constant radius of deformation with zero force, which is both physically unrealistic and mathematically disprovable.\\

To conclude, we derived a nonlinear beam equation by using Ciarlet's nonlinear plate equations \cite{Ciarlet1997theory} (see section B of Ciarlet \cite{Ciarlet1997theory}) with mathematical regiour, and show that our model is valid for constant radius of deformations.

\bibliographystyle{./model1-num-names}
\bibliography{CriticaStudyOfHowellsWork}%
\biboptions{sort&compress}

\end{document}